\documentclass[a4paper, 12pt]{article}
\usepackage[margin=2cm]{geometry}
\usepackage{amsmath}
\usepackage[hidelinks]{hyperref}
\usepackage{xcolor}
\hypersetup{colorlinks, linkcolor={red!50!black}, citecolor={blue!50!black}, urlcolor={blue!80!black}}
\usepackage{amsfonts}
\usepackage{amssymb}
\usepackage{graphicx}
\usepackage{epstopdf}
\usepackage{physics}
\begin{document}
\title{Study of geometric phase using classical coupled oscillators}
\author{Sharba Bhattacharjee, Biprateep Dey and Ashok K Mohapatra\\School of Physical Sciences, National Institute of Science Education and\\ Research, HBNI, Jatni 752050, India\\}
\date{\today}
\maketitle
\begin{abstract}
We illustrate the geometric phase associated with the cyclic dynamics of a classical system of coupled oscillators. We use an analogy between a classical coupled oscillator and a two-state quantum mechanical system to represent the evolution of the oscillator on an equivalent Hilbert space, which may be represented as a trajectory on the surface of a sphere. The cyclic evolution of the system leads to a change in phase, which consists of a dynamic phase along with an additional phase shift dependent on the geometry of the evolution. A simple experiment suitable for advanced undergraduate students is designed to study the geometric phase incurred during cyclic evolution of a coupled oscillator.
\end{abstract}
\section{Introduction}
Cyclical evolutions, where a system returns to its initial state at the end of the evolution, are often studied in Physics. The state vector of a quantum mechanical system after a cyclical evolution is related to its initial state vector by a phase factor. This phase has a component related only to the geometry of the evolution, as was shown by Pancharatnam in 1956 for cyclic evolution of the polarization state of light~\cite{pancharatnam}, and later by Berry in 1984 for adiabatic evolution of the Hamiltonian~\cite{berry}. Geometric phase for adiabatic evolution has been discussed in~\cite{griffiths}. Aharonov and Anandan generalized this to any cyclical evolution, where the geometric phase was shown to depend on the closed trajectory of the state ray in the Hilbert space~\cite{geomph}. Geometric phase has been observed experimentally in interference experiments, such as in the Aharonov-Bohm effect~\cite{aharonov-bohm}, in nuclear magnetic resonance~\cite{nmr}, and in laser optics~\cite{opt1,opt2}. This concept is being applied in a number of areas of research, such as the theory of insulators~\cite{resta} and quantum information~\cite{QI}.

In this article, the quantum mechanical system under consideration is that of a two-state atom interacting with a steadily oscillating electric field such as a laser. Such a system will exhibit Rabi oscillations~\cite{gerry,scully}, and will thus evolve cyclically. We mathematically demonstrate the motion of a system of two classical coupled oscillators to be analogous to such a system. The periodic transfer of energy between the oscillators is analogous to Rabi oscillations in the atom. This analogy is used to study the evolution of the oscillator as a trajectory on an equivalent Hilbert space and thus to obtain the geometric phase associated with the evolution. A simple experiment suitable for undergraduate students is designed to demonstrate and measure the geometric phase, and the measured value is found to be in good agreement with the expected phase. The analogy used here has been demonstrated previously in~\cite{qcanalogy}, and explored in electrical RLC circuits to demonstrate electromagnetically induced transparency (EIT)~\cite{eit}, double EIT~\cite{eit2} and Fano interference~\cite{fano}. A detailed discussion on analogies between classical and quantum mechanical systems is found in~\cite{dragoman}.  
\section{Analogy between coupled oscillator and two-level quantum system}
\begin{figure}[htb]
\centering
\includegraphics[width=\linewidth]{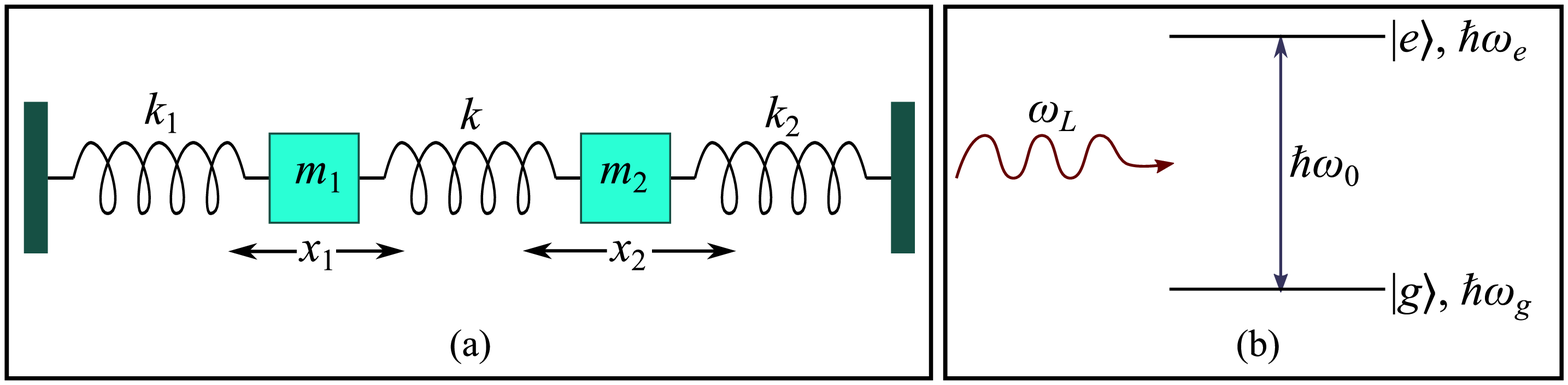}
\caption{(a) Schematic diagram of a system of coupled oscillators. (b) Level scheme of a two-state atom interacting with a single-mode laser as an example of a two-level quantum system.}
\label{fig:spma}
\end{figure}
Figure~\ref{fig:spma}(a) illustrates the classical system of two masses $m_1$ and $m_2$, attached to rigid walls with springs of spring constants $k_1$ and $k_2$ respectively and to each other with a spring $k$. If the masses are displaced by $x_1$ and $x_2$ from their respective equilibrium positions, then the equations of motion are
\begin{align}
&m_1\ddot{x}_1 = -k_1x_1+k(x_2-x_1),\\
&m_2\ddot{x}_2 = k(x_1-x_2)-k_2x_2.
\end{align}
This can be expressed as a matrix equation
\begin{equation}
-\dv[2]{t}\boldsymbol{\xi}=\mathbf{A}\boldsymbol\xi,
\label{couosc}
\end{equation}
where we have defined
\begin{equation}
\boldsymbol{\xi}=\mqty(x_1\\x_2) \qquad\text{and}\qquad \mathbf{A}=\mqty(\omega_1^2 & -\Omega_1^2 \\ -\Omega_2^2 & \omega_2^2),
\end{equation}
where $\omega_i^2=(k_i+k)/m_i$ and $\Omega_i^2=k/m_i$, $i=1,2$. $\omega_1$ and $\omega_2$ are the natural frequencies of the individual oscillators. The equations of motion can be solved in the standard manner~\cite{goldstein} to obtain the normal mode frequencies
\begin{equation} \omega_\pm^2=\frac{1}{2}\qty( {{\omega_1}}^{2} + {{\omega_2}}^{2} \pm \qty[4 {{\Omega_1}}^{2} {{\Omega_2}}^{2} + (\omega_1^2-\omega_2^2)^2]^{1/2} ). \end{equation}
The solution to \eqref{couosc} is then
\begin{equation}
\boldsymbol{\xi}=\Re[F_+\boldsymbol{\xi}_+\exp(-i\omega_+t-i\phi_+)+F_-\boldsymbol{\xi}_-\exp(-i\omega_-t-i\phi_-)], \label{eq:cosoln}
\end{equation}
where $\boldsymbol{\xi}_\pm$ are the normal mode eigenvectors. The constants $F_\pm$ and $\phi_\pm$ are determined from the initial conditions.

The system illustrated in figure~\ref{fig:spma}(b) is that of an atom with a ground state $\ket{g}$ of energy $\hbar \omega_g$ and an excited state $\ket{e}$ of energy $\hbar \omega_e$, with resonant frequency $\omega_e-\omega_g=\omega_0$. It interacts with an oscillating electric field $\mathcal{E}(t)=\Re(\mathcal{E}_0e^{i\omega_L t})$ such as a single-mode laser. We define the dipole moment \( \mu=e\mel{g}{\vb{r}\vdot \vu{\mathcal{E}}}{e} \), where $e$ is the electronic charge. Then the Hamiltonian matrix of the system in the $\qty{\ket g, \ket e}$ basis is~\cite{scully}
\begin{equation}
\mathbf{H}=\mqty(\hbar \omega_g & -\mu\mathcal{E}(t) \\ -\mu^*\mathcal{E}(t) & \hbar \omega_e).
\end{equation}
If we transform the state vector using the unitary operator \( U=\dyad{g}{g}+e^{-i\omega_L t}\dyad{e}{e} \), then the Hamiltonian matrix elements in the new basis $\qty{\ket{\tilde g}=U^\dagger \ket g =\ket g, \ket{\tilde e}=U^\dagger \ket e = e^{i\omega_L t}\ket e}$ become time independent under the rotating wave approximation, 
\begin{equation}
\mathbf{\tilde H}=\hbar\mqty( \omega_g & -{\Omega}/{2} \\ -{\Omega}/{2} & \omega_g+\Delta ), \label{eq:Hamil}
\end{equation}
where we have defined \( \Delta=\omega_0-\omega_{L} \) (called the detuning) and \( \Omega=\mu\mathcal{E}_0/\hbar \) (called the Rabi frequency). $\Omega$ is in general a complex quantity depending on the phase of the laser; here we choose the phase to be zero so that $\Omega$ is real. Physically, this may be interpreted as moving the laser source towards or away from the atom to ensure that $\mu\mathcal{E}_0$ is real. Then the state $\ket{\tilde\psi(t)}$ of the system at a time $t$, with the initial condition \( \ket{\tilde\psi(0)}=\ket{\tilde g} \), is given in the $\qty{\ket{\tilde g}, \ket{\tilde e}}$ basis as~\cite{scully} 
\begin{equation}
\ket{\tilde \psi(t)}=c_1(t)\ket{\tilde g}+c_2(t)\ket{\tilde e}\equiv \mqty(c_1(t)\\c_2(t)), \label{eq:psiexpr}
\end{equation}
where
\begin{align}
&c_1=\frac{1}{2}\left[\qty(\sqrt{\Delta^2+\Omega^2}-\Delta)\exp(-i\qty(\Delta+\sqrt{\Delta^2+\Omega^2})t/2-i\omega_gt)\right.\nonumber\\&\qquad\left.{}+\qty(\sqrt{\Delta^2+\Omega^2}+\Delta)\exp(-i\qty(\Delta-\sqrt{\Delta^2+\Omega^2})t/2-i\omega_gt)\right]\left({\Delta^2+\Omega^2}\right)^{-1/2}, \label{Rabisol1}\\
&c_2=\frac{1}{2}\left[\Omega \exp(-i\qty(\Delta-\sqrt{\Delta^2+\Omega^2})t/2-i\omega_gt)-\Omega \exp(-i\qty(\Delta+\sqrt{\Delta^2+\Omega^2})t/2-i\omega_gt)\right]\nonumber\\&\qquad\left({\Delta^2+\Omega^2}\right)^{-1/2}.
\label{Rabisol2}
\end{align}
The coefficients $c_1(t)$ and $c_2(t)$ are the probability amplitudes, so that at any given time $t$, $\abs{c_1(t)}^2$ and $\abs{c_2(t)}^2$ are the probabilities of the system collapsing to the states $\ket{\tilde g}$ and $\ket{\tilde e}$ respectively if its energy is measured.

To understand the analogy between the two systems, we look at Schr\"{o}dinger equation,
\begin{equation}
i\hbar \dv{t}\ket{\tilde \psi}=\tilde H\ket{\tilde \psi}. \label{eq:schro}
\end{equation}
For a two-dimensional Hilbert space, we can express $\ket{\tilde \psi}$ and $\tilde H$ in the basis $\qty{\ket{\tilde g}, \ket{\tilde e}}$ to write \eqref{eq:schro} as a matrix equation. Differentiating \eqref{eq:schro} with respect to $t$, we obtain
\begin{equation}
-\dv[2]{t}\ket{\tilde \psi}=\hbar^{-2}\tilde H^2\ket{\tilde \psi}, \label{eq:schro2}
\end{equation}
since the Hamiltonian matrix elements are independent of time in the chosen basis. If the state vector $\ket{\tilde \psi}$ is taken to be analogous to the displacement vector $\boldsymbol\xi$ of the coupled oscillator, then~\eqref{eq:schro2} is seen to be equivalent to \eqref{couosc} if we have
\begin{equation}
\mathbf{A}=\hbar^{-2}\mathbf{\tilde H}^2, \label{eq:reln}
\end{equation}
where $\mathbf{\tilde H}$ is the matrix of the Hamiltonian in the chosen basis. Thus, as long as $\mathbf{A}$ is hermitian (symmetric in this case, since its entries are real), the coupled oscillator system can be found to be analogous to a two-state quantum system whose Hamiltonian is obtained from \eqref{eq:reln}. The displacements $x_1$ and $x_2$ observed from the experiment are then analogous to the real parts of the coefficients $c_1$ and $c_2$ in \eqref{eq:psiexpr}, up to normalization. It should be noted that  $\ket{\tilde\psi}$ and $\boldsymbol\xi$ are physically different quantities, even though they evolve with time following mathematically similar equations. The quantities $\Omega$ and $\Delta$ have no exact physical meaning for the coupled oscillator, but they roughly indicate the coupling strength and the difference between the natural frequencies respectively. The relation between the quantities associated with the two systems has been shown in table~\ref{tab:qtyreln}.
\begin{table}[ht]
\caption{Relation between the quantities associated with the classical coupled oscillator and the equivalent quantum two-level system.}
\label{tab:qtyreln}
\begin{tabular}{ll}
\hline
Coupled oscillator & Two-level system\\
\hline
Oscillator displacements $x_1$, $x_2$ & Real parts of probability amplitudes $c_1$, $c_2$\\
First natural frequency $\omega_1$ & $(\omega_g^2+\Omega^2/4)^{1/2}$\\
Second natural frequency $\omega_2$ & $[(\omega_g+\Delta)^2+\Omega^2/4]^{1/2}$\\
Coupling $\Omega_1=\Omega_2$ & $[\Omega(2\omega_g+\Delta)/2]^{1/2}$\\
\hline
\end{tabular}
\end{table}
\section{Experimental observation of normal mode frequencies}
\begin{figure}[!h]
\centering
\includegraphics[width=\linewidth]{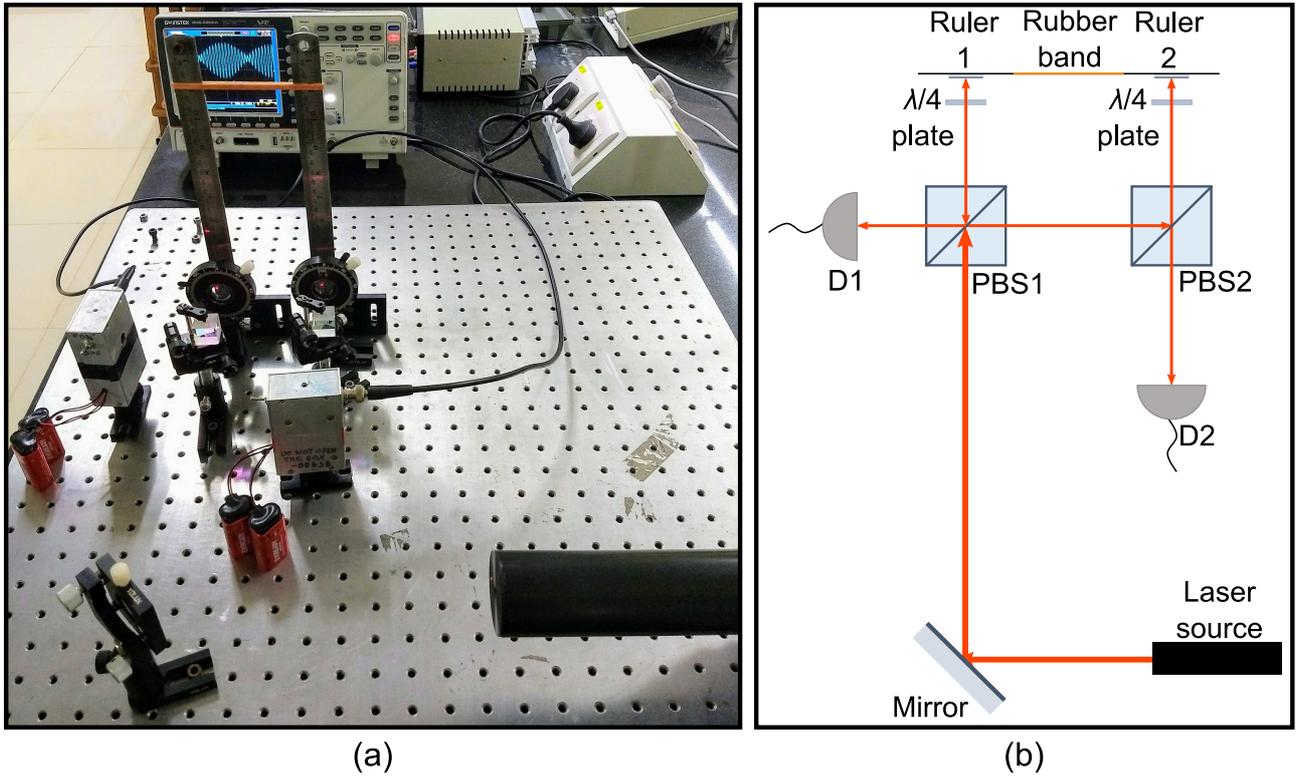}
\caption{(a) Photograph of the experimental set-up. The two vertical metal rulers connected by a rubber band serve as a system of coupled oscillators. Their motion is studied using a laser beam as explained in the text. (b) Schematic diagram of the experimental set-up (top view) showing the path taken by the laser beam. PBS1, PBS2: polarizing cube beam splitters; D1, D2: quad photodetectors.}
\label{fig:setup_both}
\end{figure}
The experiment is set up as shown in figure~\ref{fig:setup_both}. Two vertical vibrating metal rulers fixed at the bottom serve as the oscillating bodies, which are coupled by connecting them with a rubber band. The motion of this set-up is similar to that of the coupled oscillator described in the previous section, where the masses and spring constants of the equivalent spring-mass system depend on the masses and elastic moduli of the rulers and on the position and elastic properties of the rubber band.


The motion of the rulers can be observed using cameras or any kind of motion sensors. In our case, a He-Ne laser beam is used to accurately measure the displacements of the rulers. The unpolarized beam is split into two polarized beams with perpendicular planes of polarization using the polarizing cube beam splitter PBS1. The transmitted beam is made to reflect off a plane mirror mounted on the first ruler. It is also made to pass through a quarter-wave plate aligned to rotate the plane of polarization of the beam by $\pi/2$, so that it is then reflected by PBS1 onto the quad photodetector D1 instead of being transmitted back towards the laser source. The optical arrangement is shown schematically in figure~\ref{fig:setup_both}(b).
\begin{figure}[h!]
\centering
\includegraphics[width=0.7\linewidth]{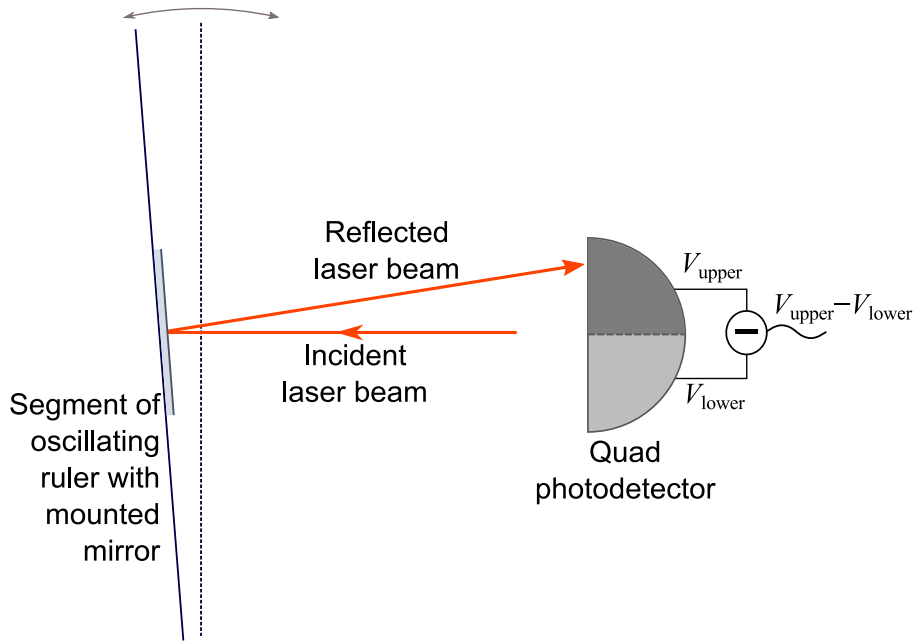}
\caption{Schematic diagram of a portion of the experimental set-up (side view) showing the relation between the vibration of the ruler and the oscillation of the laser spot incident on the quad photodetector. The beam splitters and wave plates have not been shown in this figure for simplicity.}
\label{fig:Oscill}
\end{figure}

When the the ruler vibrates, the small segment of it to which the mirror is attached describes angular oscillation about the vertical, which can be considered to be simple harmonic as long as the amplitude of motion is small. The laser beam reflected off the mirror thus describes angular oscillation about the horizontal, and the laser spot on the detector moves up and down with its motion proportional to that of the ruler as shown in figure~\ref{fig:Oscill}. A quad photodetector is used, and the signals detected by the upper and lower halves of the detector are separately amplified (with the same gain) and subtracted. As the laser spot moves upward from the mean position, the upper half of the photodetector detects a larger signal than the lower half, and the reverse happens when it moves downward. Therefore, the difference between the output voltages of the two halves is proportional to the displacement of the oscillating ruler, and this is measured using an oscilloscope. The photodetector circuit has been further described in the appendix. The amplitude of oscillation has no significance for demonstrating the analogy between the oscillator and the two-level atom, and we normalize the signals of the two oscillators so that the sum of the squares of their amplitudes is initially $1$, calling them the normalized displacements. Similarly, the beam reflected by PBS1 is further reflected by a second polarizing cube beam splitter PBS2 and then by a plane mirror mounted on the second ruler. This beam is also made to pass through a quarter wave plate to rotate its plane of polarization so that it is is then transmitted by PBS2 onto the quad photodetector D2, where the signal is recorded in a similar manner.
\begin{figure}[!h]
\centering
\includegraphics[width=0.9\linewidth]{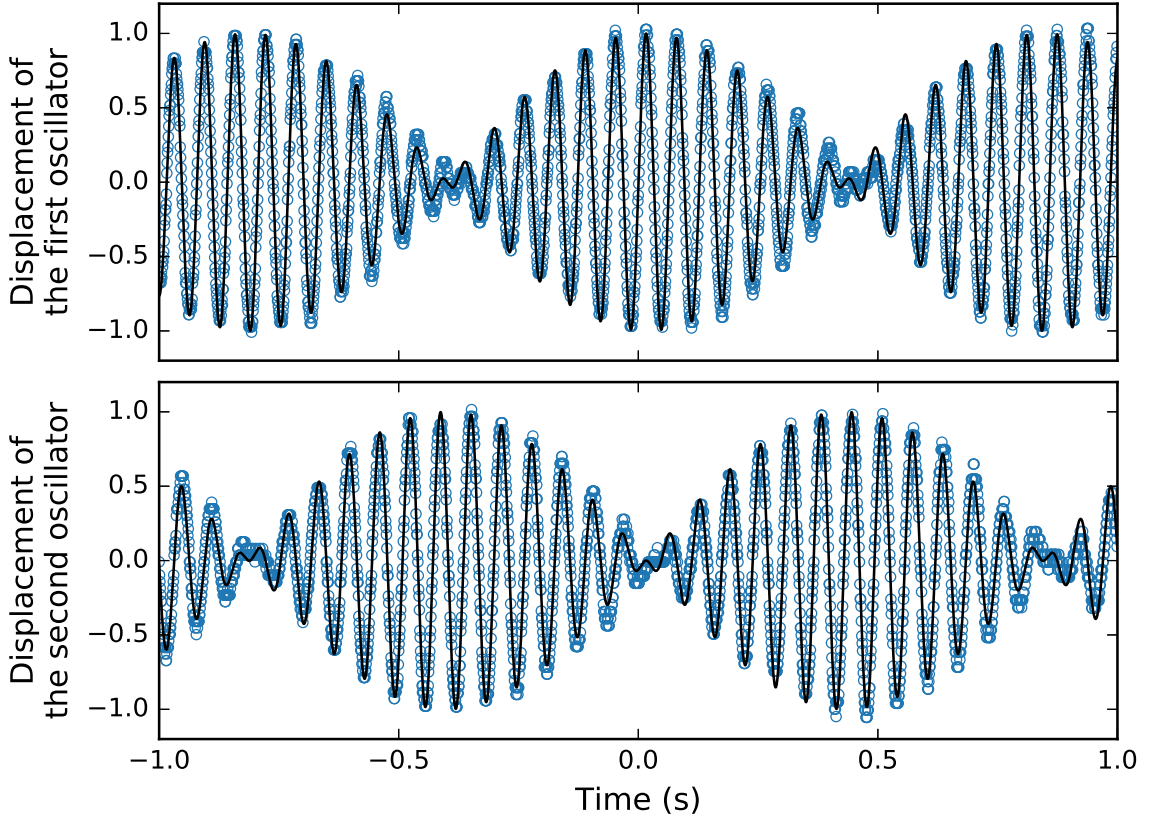}
\caption{Normalized displacements of the two oscillators as functions of time $t$. The open circles are the experimental data, and the solid lines denote the fitted functions. The fitted parameters for this plot are $\omega_g=98.917$~rad/s, $\Delta=-0.088$~rad/s and $\Omega=7.560$~rad/s.}
\label{fig:sig2}
\end{figure}

\begin{figure}[h!]
\centering
\includegraphics[width=0.75\linewidth]{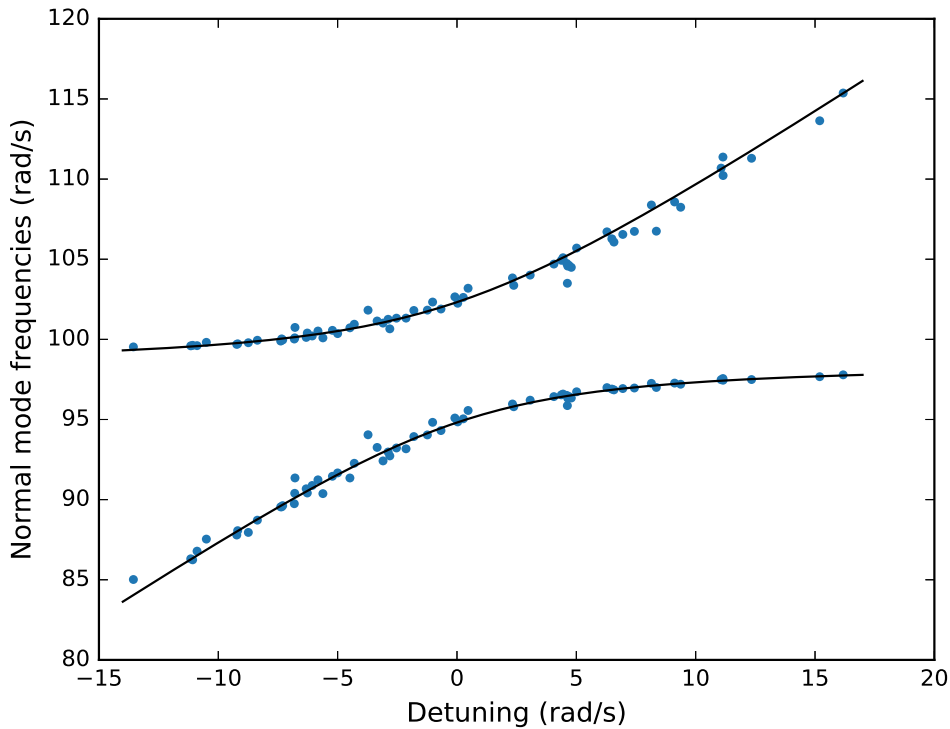}
\caption{Normal mode frequencies ($\omega_\pm$) as functions of the detuning $\Delta$. The circles are the experimental data, and the solid lines denote the fitted curves. The parameters determined from the fitting are $\omega_g=98.44~\text{rad/s}$ and $\Omega=7.82~\text{rad/s}$.}
\label{fig:3delvsfreqs}
\end{figure}

We strike one of the rulers to start the oscillation and record the time-evolution of the normalized oscillator displacements as shown in figure~\ref{fig:sig2}. It is seen that both the oscillator displacements have beat-like forms, exhibiting rapid oscillations whose amplitudes are also varying periodically with a relatively longer period. The rapid oscillations correspond to the oscillatory motions of the individual oscillators, while the periodic variation in amplitude is caused by the transfer of energy between the two oscillators due to the coupling. For the equivalent quantum mechanical system, the rapid oscillations are interpreted as the result of the dynamic phase incurred by the state vector, while the periodic variation of amplitudes is indicative of Rabi oscillations. The displacement of the ruler that was initially struck is fitted with the function
\begin{equation} f(t)=A\cos[\omega_+ (t-d)+\phi]+B\cos[\omega_- (t-d)+\phi]+C. \label{eq:fitfn} \end{equation}
Damping is ignored in this fitting, since the damping of the oscillation over the recording period of 2 seconds is insignificant. The fitted values of $A$, $B$, $\omega_+$ and $\omega_-$ are used to calculate the parameters $\omega_g$, $\Delta$ and $\Omega$ of Rabi model by comparison with \eqref{Rabisol1}, and are obtained as
\begin{equation}
\Delta=(\omega_+-\omega_-)\frac{B-A}{B+A},\qquad \Omega=(\omega_+-\omega_-)\qty[1-\qty(\frac{B-A}{B+A})^2]^{1/2},\qquad\omega_g=\frac{\omega_++\omega_--\Delta}{2}.
\end{equation}
$C$ and $d$ are introduced to account for offsets, and the global phase factor $\phi$ is set equal to zero. We change the effective detuning $\Delta$ by attaching small masses to one of the rulers to change the natural frequency (this also changes the other parameters $\omega_g$ and $\Omega$, but by a relatively smaller amount), and thus obtain the parameters for different values of detuning. The normal mode frequencies $\omega_{\pm}$ determined from the fitting have been plotted as functions of the effective detuning $\Delta$ in figure~\ref{fig:3delvsfreqs}.
\section{Phase change and geometric phase in coupled oscillator}
The beat-like evolution of the oscillator displacements can be explained by noting from Eqs.~\eqref{eq:cosoln}, \eqref{Rabisol1} and \eqref{Rabisol2} that the coupled oscillator displacements as well as the state vector coefficients for the two-state atom evolve with time in the form \begin{equation} f(t)=Ae^{-i\omega_+t}+Be^{-i\omega_-t}. \end{equation} Here, $f(t)$ can be considered a coefficient of the atomic state vector, or we can equivalently take $\Re(f(t))$ to be one of the coupled oscillator coordinates. Then we have \begin{equation} |f|^2=|A|^2+|B|^2+2AB\cos[(\omega_+-\omega_-)t], \end{equation} which indicates an oscillatory behavior with a frequency of $\Omega_R=\omega_+-\omega_-$. The graph of $\Re(f(t))$ as a function of $t$ is thus expected to be sinusoidal, where the envelope of the sinusoid is periodic with frequency $\omega_+-\omega_-$, as can be seen in figure~\ref{fig:sig2}. For the two-state atom, $\Omega_R$ is called the effective Rabi frequency, and the system evolves cyclically, returning to its initial state after one Rabi period $T_R=2\pi/(\omega_+-\omega_-)$. However, the state vector can incur a phase shift during the Rabi oscillation, which may include a geometric phase.

To find the phase change, we note that $f(0)=A+B$, but $f(2\pi/(\omega_+-\omega_-))=(A+B)\exp[-2\pi i\omega_-/(\omega_+-\omega_-)]$. The global phase change is thus $-2\pi\omega_-/(\omega_+-\omega_-)$. From Eqs.~\eqref{Rabisol1} and \eqref{Rabisol2}, we see that the effective Rabi frequency for a two-state atom is $\Omega_R=\sqrt{\Delta^2+\Omega^2}$, and the phase shift after one Rabi period is
\begin{equation} \phi=\frac{-2\pi \omega_g}{\sqrt{\Delta^2+\Omega^2}}+\pi\left(1-\frac{\Delta}{\sqrt{\Delta^2+\Omega^2}}\right). \end{equation}
The phase change thus has two components. The state of the system has a constant average energy of $\hbar\omega_g$, and the first term of the phase shift is simply the corresponding frequency $\omega_g$ multiplied by the time period \( T_R=2\pi/\sqrt{\Delta^2+\Omega^2} \) for one Rabi oscillation. This is the dynamic phase change of the system. The second term depends only on the parameters $\Delta$ and $\Omega$, and not on the energy $\omega_g$. This phase change is the geometric phase associated with the quantum evolution. The geometric phase $\phi_G$ is thus given by
\begin{equation}
\phi_G=\pi\left( 1-\frac{\Delta}{\sqrt{\Delta^2+\Omega^2}} \right). \label{eq:geomph}
\end{equation}
At resonance ($\Delta=0$), we have $\phi_G=\pi$. For very high detuning ($\Delta\to\pm\infty$), we get $\phi_G\to 0$, noting that phase is unique only up to an added multiple of $2\pi$.
\begin{figure}[!h]
\centering
\includegraphics[width=0.65\linewidth]{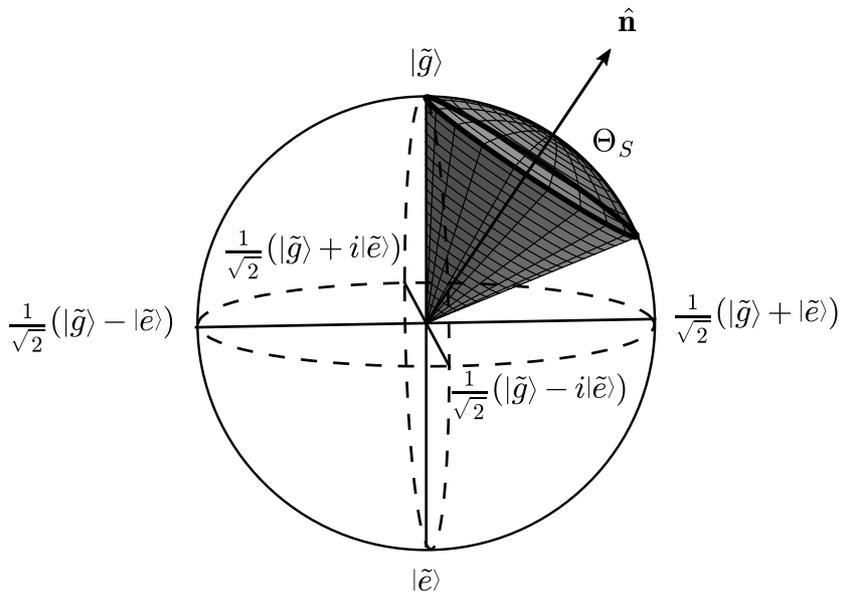}
\caption{Trajectory of the Bloch vector for $\Omega=0.8$ and $\Delta=1.2$, with $\vu{n}$ showing the direction of the fictitious magnetic field. The area of the shaded spherical cap gives the solid angle $\Theta_S$ subtended by the trajectory.}
\label{fig:bloch2}
\end{figure}

The geometric interpretation of $\phi_G$ can be seen from the Bloch sphere representation of the system. We know that any state of a two-level quantum mechanical system can be represented as a unit vector, called the Bloch vector~\cite{gerry,2ststuff}. The Bloch vector for the initial state $\ket{\tilde g}$ is the unit vector along the positive $z$-axis. The Hamiltonian obtained in \eqref{eq:Hamil} is seen to be
\begin{equation}
\tilde H=-\hbar\left[ \left(-\omega_g-\frac{\Delta}{2}\right) I+\Omega  S_x+\Delta  S_z \right],
\end{equation}
where $I$ is the identity operator, and $S_i=\sigma_i/2\ (i=x,y,z)$ are the generators of rotation about the axes. $\sigma_i$, expressed in the $\{ \ket{\tilde g},\ket{\tilde e} \}$ basis, are the Pauli matrices. This is equivalent to the dynamics of a spin-$1/2$ particle in presence of a fictitious magnetic field $\vb{B}=(\Omega\vu{e}_x+\Delta \vu{e}_z)$. The Bloch vector will rotate about $\vb{B}$, with its tip tracing out a circle. The solid angle subtended by this circle at the center of the Bloch sphere is \( \Theta_S=2\pi\left( 1-\Delta/\sqrt{\Delta^2+\Omega^2} \right) \). The geometric phase is half of this solid angle~\cite{geomph}, which gives us the expression obtained in \eqref{eq:geomph}. The trajectory of the Bloch vector is shown in figure~\ref{fig:bloch2}.
\section{Observation of geometric phase in coupled oscillator}
\begin{figure}[!h]
\centering
\includegraphics[width=0.9\linewidth]{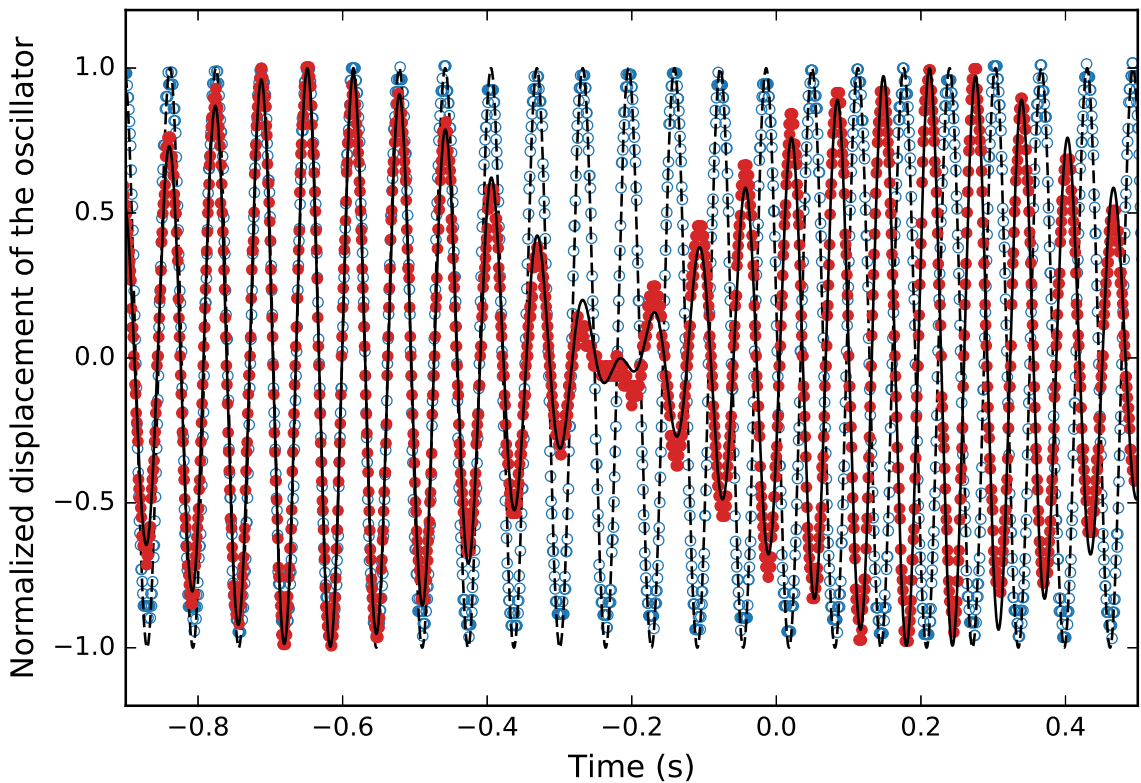}
\caption{Plot showing geometric phase in a coupled oscillator for \( \Delta=0.030~\text{rad/s},\) and \(\Omega=7.402~\text{rad/s}\). The open circles represent the data for the uncoupled oscillator, fitted with the dashed line. The closed circles represent the data for the coupled oscillator with small detuning, fitted with the solid line.}
\label{fig:cpvsun}
\end{figure}
Figure~\ref{fig:cpvsun} illustrates geometric phase in a coupled oscillator by comparing the time-evolution of displacement of one oscillator for low detuning with that of the same oscillator for high detuning (which is equivalent to an uncoupled system). At $t=0$, the phases of the displacements of the coupled and uncoupled oscillators are made to match. The frequency of the uncoupled oscillator is approximately $\omega_g$, and thus the phase change of its displacement after one Rabi period is roughly equal to the dynamic phase of the coupled oscillator. The phase difference between the two signals after one Rabi period thus gives an estimate of the geometric phase incurred by the coupled oscillator after a cyclic evolution. From figure~\ref{fig:cpvsun}, the phase change is seen to be close to $\pi$, as expected for a system with low detuning. From the fitted parameters \( \Delta=0.030~\text{rad/s}\) and \(\Omega=7.402~\text{rad/s}\), the geometric phase is calculated to be $0.996\pi$.
\begin{figure}[h!]
\centering
\includegraphics[width=0.9\linewidth]{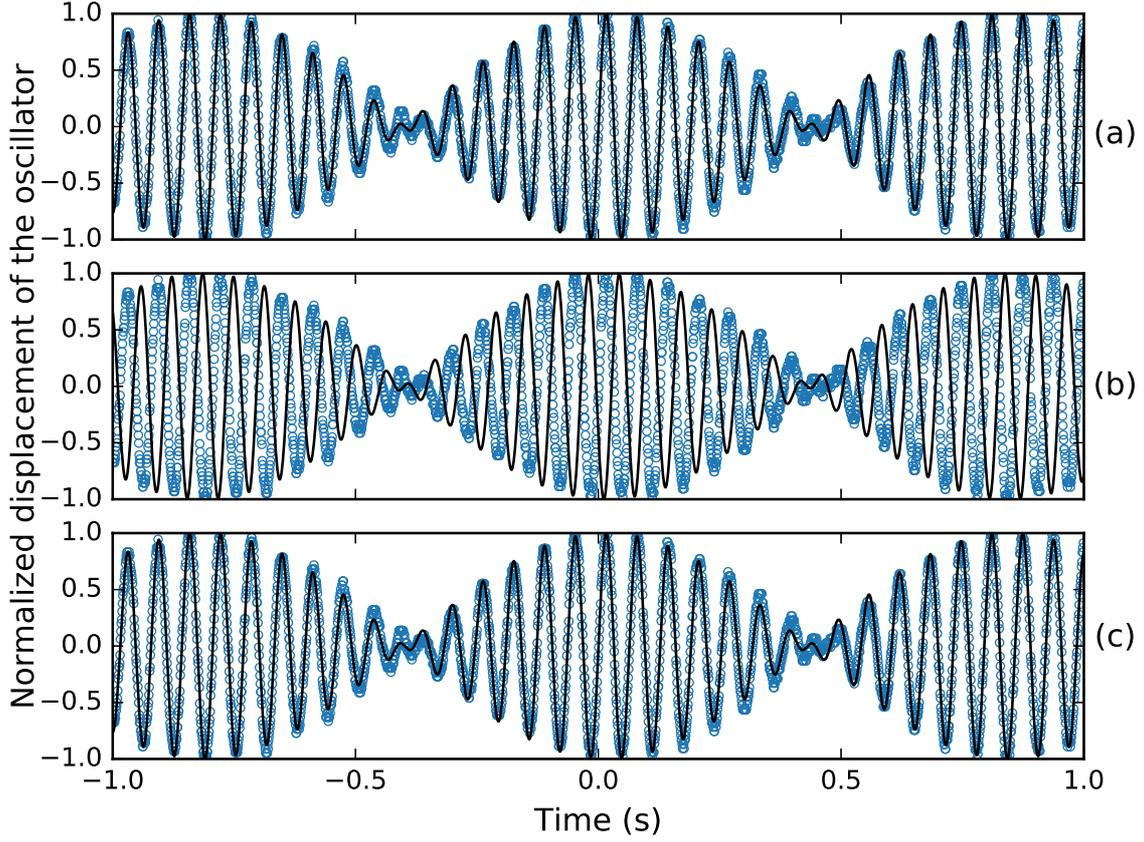}
\caption{Procedure to determine the phase shift. In each of the three graphs, the open circles represent the data for the same oscillator signal. In (a), the graph shows the function fitted to the data, represented by the solid line. In (b), the function has been shifted by one Rabi period. The envelopes match, but a phase difference remains between the data and the plotted function. In (c), the phase has been fitted to match the function with the data keeping the other parameters fixed, and this gives the total phase shift.}
\label{fig:plots}
\end{figure}
\begin{figure}[!h]
\centering
\includegraphics[width=0.75\linewidth]{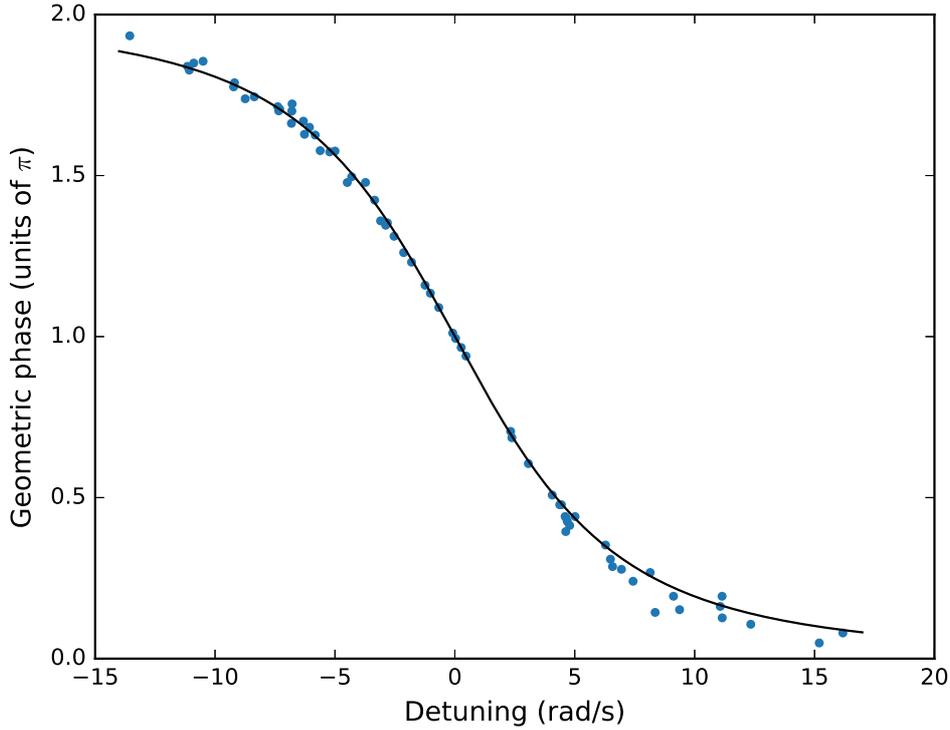}
\caption{Geometric phase $\phi_G$ as a function of the detuning $\Delta$. The circles are the experimental data, and the solid line denotes the fitted curve. The parameter determined from the fitting is $\Omega=7.32~\text{rad/s}$.}
\label{fig:3delvsphi}
\end{figure}

To measure the phase change, the effective Rabi frequency is at first calculated from the fitted parameters. Then the fitted function is shifted along the $t$-axis by one Rabi period (by replacing $d$ with $d-2\pi/(\omega_+-\omega_-)$ in $f(t)$). The phase parameter $\phi$ in \eqref{eq:fitfn} is then adjusted to obtain the total phase change over one Rabi period. From this, the dynamic phase of $-2\pi \omega_g/(\omega_+-\omega_-)$ is subtracted to obtain the geometric phase, which is compared with the geometric phase calculated from the fitted values of $\Delta$ and $\Omega$ using \eqref{eq:geomph}. The method for obtaining the phase shift is demonstrated in figure~\ref{fig:plots}. The geometric phase corresponding to different values of detuning were determined using the above analysis, which have been shown in figure~\ref{fig:3delvsphi}. The geometric phase in the coupled oscillator as measured from the experiment is found to be in very good agreement with the expected phase based on the theory of a two-state atom.
\newpage
\section{Conclusions}
We have demonstrated that the classical system of coupled oscillators evolves in a manner similar to a two-state atom in a steadily oscillating electric field. This analogy may be used to simulate in classical coupled oscillators various phenomena associated with the dynamics of quantum mechanical systems. The Rabi model parameters of the equivalent two-level quantum mechanical system have been calculated, and the geometric phases associated with the cyclical evolution have been estimated and found to be in very good agreement with the expected values. This experiment can be used to get a better insight of the concept of geometric phase, which usually does not have any simple demonstration suitable for undergraduate students. The experiment described here can be easily reproduced in undergraduate laboratories with commonly available equipment.

\section*{Acknowledgments}
The authors are grateful to the Department of Atomic Energy, Government of India for providing with the necessary financial support, to Dr.\ Yogesh Kumar Srivastava and Dr.\ Joydeep Bhattacharjee for their insightful suggestions, and to Mr.\ Rudranarayan Mohanty for his technical assistance in setting up the experiment. The authors acknowledge Mr.\ Evan John Philip and Mr.\ Kunal Garg for their lab report on the demonstration of the analogy between a coupled oscillator and a two-state quantum system.
\newpage
\appendix
\section*{Appendix: Photodetector design}
\setcounter{section}{1}
\begin{figure}[h!]
\centering
\includegraphics[width=0.7\linewidth]{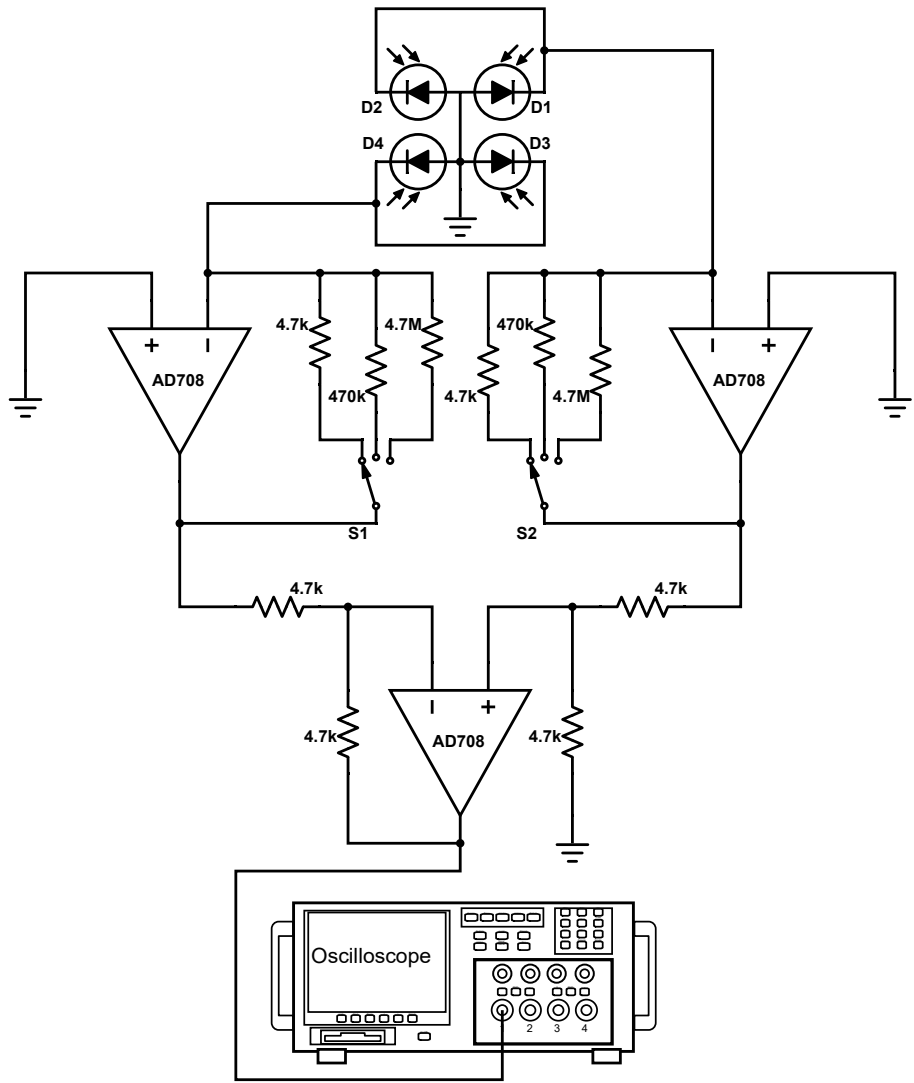}
\caption{Circuit diagram of a photodetector. The quad photodetector consists of four photodiodes arranged in a square. The upper two photodiodes together detect the upper laser signal, while the lower two together detect the lower signal. Their outputs are separately amplified with the same gain using AD708 op-amps, and subtracted using a differential amplifier to obtain the signal voltage corresponding to the displacement of the ruler, which is observed in an oscilloscope.}
\label{fig:ckt}
\end{figure}

The photodetector circuit diagram is shown in figure~\ref{fig:ckt}. Quad photodetectors were used, and the voltages in the upper and lower photodiodes were separately amplified and then subtracted. The circuit was soldered onto a prototyping board. The $\pm V_{CC}$ voltages were supplied using $9$~V batteries. The outputs were fed to oscilloscopes using BNC cables.

\bibliography{couposcbib}
\bibliographystyle{unsrt}
\end{document}